\documentclass[epj]{svjour}
\usepackage{latexsym}
\usepackage{graphics}

\def\br{\begin{eqnarray}}
\def\er{\end{eqnarray}}
\def\be{\begin{equation}}
\def\ee{\end{equation}}

\def\g{\gamma}

\def\m{\mu}

\def\({\left(}
\def\){\right)}
\def\<{\left\langle}
\def\>{\right\rangle}
\def\gsim{\stackrel{\scriptstyle >}{\phantom{}_{\sim}}}

\def\S{\Sigma}

\begin{document}

\title{The origin of the first and third generation fermion masses in a technicolor scenario}

\author{ A. Doff and A. A. Natale}

\institute{Instituto de F\'{\i}sica Te\'orica, UNESP, Rua Pamplona 145, 01405-900, S\~ao Paulo, SP, Brazil}
\date{Received: date / Revised version: date}

\abstract{ We argue that the masses of the first and third fermionic generations, which are respectively of the
order of a few MeV up to a hundred GeV, are originated in a dynamical symmetry breaking mechanism leading to
masses of the order $\alpha \mu$, where $\alpha$ is a small coupling constant and $\mu$, in the case of the
first fermionic generation, is the scale of the dynamical quark mass ($\approx 250$ MeV). For the third fermion
generation $\mu$ is the value of the dynamical techniquark mass ($\approx 250$ GeV). We discuss how this
possibility can be implemented in a technicolor scenario, and how the mass of the intermediate generation is
generated. 
}

\maketitle
\section{Introduction}
\par The standard model is in excellent agreement with the experimental data. The only still obscure part of
the model is the one responsible for the mass generation, i.e. the Higgs mechanism. In order to make the
mass generation mechanism more natural there are several alternatives, where the most popular ones are supersymmetry
and technicolor. In the first one the mass generation occurs through the existence of non-trivial
vacuum expectation values of fundamental scalar bosons while in the second case the bosons responsible for
the breaking of gauge and chiral symmetry are composite. Up to now the fermionic mass spectrum is the
strongest hint that we have in order to unravel the symmetry breaking mechanism.
A simple and interesting way to describe the fermionic mass spectrum is to suppose that the mechanism behind
mass generation is able to produce a non-diagonal mass matrix with the Fritzsch texture \cite{fritzsch}
\br
 M_f =\left(\begin{array}{ccc} 0 & A & 0\\ A^* & 0 & B \\
0 & B^* & C
\end{array}\right).
\label{e1} \er
This matrix is similar for the charged leptons, $1/3$ and $2/3$ charged quarks. The entry $C$ is proportional
to the mass of the third generation fermion, while the entry $A$ is proportional to the mass
of the lighter first generation. The diagonalization of such mass matrix will determine the
CKM mixing angles and the resulting diagonal mass matrix should reproduce the observed current
fermion masses. There are other possible patterns for the mass matrix and we choose the one of
Eq.(\ref{e1}) just for simplicity.  We call attention to the values of $A$ and $C$. They must be of
order of a few MeV and a hundred GeV respectively. In models with a fundamental Higgs boson
the values of $A$ and $C$ are obtained due to adjusted vacuum expectation values (vev) or Yukawa
couplings. In this way there is no natural explanation for the values of  $A$ and $C$; they appear
just as an {\sl ad hoc} choice of couplings!
The question that we would like to discuss here is how we naturally can generate the scales  $A$ and $C$?
In order to do so let us recall which are the mass scales in the standard model. In this model
we have basically two natural mass scales:  $\m_{qcd} \approx 250$ MeV, which is the quantum chromodynamics (qcd)
dynamical quark mass scale and $v \approx 250$ GeV, the vacuum expectation value of the fundamental Higgs field responsible for
the gauge symmetry breaking. As qcd is already an example of a theory with dynamical symmetry
breaking we will also assume that technicolor (tc) models provide a more natural way
to explain the gauge symmetry breaking \cite{weisus,hillsi}, {\it i.e.} at this level all the symmetry breaking
mechanisms should be dynamical. Therefore we will not discuss about a fundamental scalar
field with  vev $v$ but of a composite scalar field characterized by $\m_{tc} \approx 250$ GeV,  which is the
scale of the dynamical techniquark mass. Of course, at very high energies we possibly have other natural
mass scales as the Planck one, a grand unified theory (gut) scale $M_{gut}$  or a horizontal (family)
symmetry mass scale $M_h$, although it is far from clear how such scales interfere with the values
of $A$ and $C$. Finally, in tc models we may also have the extended technicolor (etc) mass scale
$M_{etc}$ \cite{dim} upon
which no constraint can be established above the $1$ TeV scale \cite{will}.
In this work we will build a model where the scales  $A$ and $C$ of Eq.(\ref{e1}) can be
related respectively to the scales  $\m_{qcd}$  and  $\m_{tc}$ times some small coupling constant. The values
of Eq.(\ref{e1}) will depend the least as possible on the very high energy mass scales like $M_ {gut}$,  $M_{etc}$, etc ...
The model  will require a very peculiar dynamics for the tc theory as well as for qcd, and this  peculiarity in what
concerns qcd differs the present approach from any other that may be found in the literature. In the next
section we discuss which is the dynamics of non-Abelian theories that will lead to the desired relation
between $A(C)$ and $\m_{qcd}(\m_{tc})$. In Section III we introduce a model assuming that its strongly
interacting sector has the properties described in the previous section, and show that the intermediate
mass scale ($B$) of Eq.(\ref{e1}) appears naturally in such a scheme. In Section IV we compute the fermion mass
matrix. Section V contain some brief comments about the pseudo goldstone bosons that appear in our model and
we draw our conclusions in the last section.
\section{The self-energy of quarks and techniquarks }
\begin{figure}[htb]
\begin{center}
\includegraphics{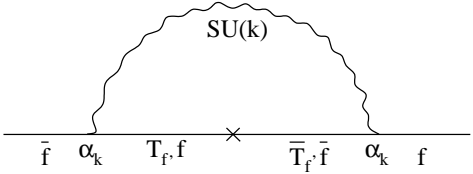} \vspace{0.5cm} \caption{Typical diagram contributing to the fermion masses of the
first and third generation. The exchange of the boson indicated by $SU(k)$ plays the same role of an extended
technicolor boson.} \label{diagmassa}
\end{center}
\end{figure}

In tc models the ordinary fermion mass is generated through the diagram shown in Fig.(\ref{diagmassa}). In
Fig.(\ref{diagmassa}) the boson indicated by  $SU(k)$
corresponds to  the exchange of a non-Abelian boson, with  coupling $\alpha_k$ to fermions
($f$) or technifermions ($T$). In the models found in the literature the role of the $SU(k)$ group
is performed by the extended technicolor
group and the boson mass is given by $M_{etc}$
To perform the calculation of Fig.(\ref{diagmassa}) we can use the following general expression for the techniquark (or
quark)  self-energy \cite{dn}
 \be \S(p)_{g} = \mu\(\frac{\mu^2}{p^2}\)^{\theta}[1 +
bg^2_{tc(qcd)}(\mu^2)ln(p^2/\mu^2)]{,}^{\!\!\!-\gamma{\cos(\theta\pi)}} \label{mger1} \ee
\noindent where in the last equation we identified $\gamma = \gamma_{tc(qcd)}$ as the canonical anomalous
dimension of the tc(qcd) mass operator, and $\mu$ is the dynamical fermion (tc or qcd) mass.  The advantage of using such
expression is that it interpolates between the extreme possibilities for the technifermion (or quark) self-energy, {\it i.e.}
 when $\theta=1$ we have the soft self-energy giving by
\be \hspace{0.5cm}\S_s (p) = \frac{\mu^3}{p^2}[1 + bg_{tc(qcd)}^2(\mu^2)ln(p^2/\mu^2)]^{\g}, \label{eq2} \ee
which is the one obtained when the composite operator  $\< \bar{\psi_{i}}\psi_{i} \> \equiv \mu^3_i$ has
canonical dimension and where $i$ can indicate $tc$ or $qcd$. When $\theta=0$ operators of higher dimension may
lead to the hard self-energy expression \be \S_h (p) = \mu[1 + bg^2_{tc(qcd)}(\mu^2)ln(p^2/\mu^2)]^{-\g},
\label{einc} \ee where $\g$ must be larger than $1/2$ and the self-energy behaves like a bare mass \cite{kl}.
Therefore no matter is the dimensionality of the operators responsible for the mass generation in technicolor
theories the self-energy can always be described by Eq.(\ref{mger1}). In the above equations $g_{tc(qcd)}$ is
the technicolor(qcd) coupling constant and $\g =\frac{3c_{tc(qcd)}}{16\pi^2b}$, where
$c_{tc(qcd)}=\frac{1}{2}[C_2(R_{1}) + C_2(R_{2}) - C_2(R_{\overline{\psi}\psi})]$ , with the quadratic Casimir
operators $C_2(R_{1})$ and $C_2(R_{2})$ associated to the $R.H$ and $L.H$ fermionic representations of the
technicolor(qcd) group, and $C_2(R_{\overline{\psi}\psi})$ is related to the condensate representation. $b$  is
the $g_{tc(qcd)}^3$ coefficient of the technicolor(qcd) group $\beta$ function. The complete equation for the
dynamical fermion mass displayed in Fig.(\ref{diagmassa}) is \be
 m_{f} = a_{k}\int dq^4\left(\frac{\mu^2}{q^2}\right)^{\theta}\frac{g^{2}_{k}(q)[1
+ b_{tc(qcd)}g^2_{tc(qcd)}ln(\frac{q^2}{\mu^2})]^{-\delta}}{(q^2 + M^{2}_{k})(q^2 +
\mu^2_{tc(qcd)})},\label{eqmf} \ee where we define $a_{k}= \frac{3C_{2k}\mu_{tc(qcd)}}{{16\pi^4}}$. In the last
equation $C_{2k}$ is the Casimir operator related to the fermionic representations of the $SU(k)$ (or etc) group
connecting the different fermions (tc or qcd), $g_k$ and $M_k$ are the respective coupling constant and gauge
boson mass, a factor $\mu_{tc(qcd)}$ remained in the fermion propagator as a natural infrared regulator and
$\delta =\gamma{\cos\theta\pi}$, $g^{2}_{k}(q)$ is assumed to be giving by \be g^2_{k}(q^2) \simeq
\frac{g_{k}^2(M^2_{k})}{(1 + b_{k}g^2_{k}(M^2_{k})ln(\frac{q^2}{M^2_{k}}))} . \label{eqgetc} \ee Note that in
Eq.(\ref{eqmf})  we have two terms of the form $[1+b_i g^2_i ln q^2]$ where the index $i$ can be related to
tc(qcd) or $SU(k)$. To obtain an analytical formula for the fermion mass we will consider  the substitution $q^2
\rightarrow \frac{xM^2_{k}}{\m^2_i}$, and we will  assume that $b_{k}g^2_{k}(M_{k}) \approx
b_{tc(qcd)}g^2_{tc(qcd)}(M_{k})$ , what will simplify considerably the calculation. Knowing that the $SU(k)$
group usually is larger than the tc(qcd) one, we computed numerically the error in this approximation for few
examples found in the literature. The resulting expression for $m_f$ will be overestimated by a factor $1.1 -
1.3$ and is giving by
 \be m_{f}\simeq\frac{3C_{2k}g^{2}_{k}(M_{k})\mu}{16\pi^2}\left(\frac{\mu^2}{M^2_{k}}\right)^{\!\!\!\theta}\!\!\!\left[1 +
b_{tc(qcd)}g^2_{tc(qcd)}ln\frac{M^2_{k}}{\m^2}\right]^{-\delta}\!\!\!\!\!\!I, \label{eqfin}
 \ee
 where
$$
I = \frac{1}{\Gamma(\sigma)}\int_{0}^{\infty}\!\!\!d{\sigma}\sigma^{\epsilon -1}e^{-\sigma}\frac{1}{\theta +
\rho\sigma}.
$$
with $\rho=b_{tc(qcd)}g^2_{tc(qcd)}(M_{k})$ and  $\epsilon = \delta + 1 = \gamma {cos\theta\pi} + 1$. To obtain
Eq.(\ref{eqfin}) we made use of the following Mellin transform
 \be \left[ 1 + \kappa \ln \frac{x}{\mu^2} \right]^{-\epsilon} \!\!\!\!=
\frac{1}{\Gamma(\epsilon)}\int_0^\infty d\sigma \, e^{-\sigma} \left( \frac{x}{\mu^2} \right)^{-\sigma
\kappa}\!\!\!\!\!\! \sigma^{\epsilon - 1}. \label{eqx} \ee \noindent Finally,  we obtain
 \be m_{f} \simeq\ \,\frac{3C_{2k}g^{2}_{k}(M_{etc})\mu}{16\pi^2}\left(\frac{\mu^2}{M^2_{etc}}\right)^{\theta}F(\cos\theta\pi,\gamma,\rho).
 \label{eqfin2}\ee
\noindent where \br F(\cos\theta\pi,\gamma,\rho) = &&\Gamma(-\gamma \cos(\theta\pi),\frac{\theta}{\rho})
\exp({\frac{\theta}{\rho}})\frac{1}{\rho}\left(\frac{\theta}{\rho}\right)^{\gamma \cos(\theta\pi)}\nonumber \\
&&\left[1 + b_{tc(qcd)}g^2_{tc(qcd)}ln\frac{M^2_{k}}{\m^2}\right]^{-\gamma {\cos(\theta\pi)}}\hspace{-1.2cm}.
\nonumber\er
\par Simple inspection of the above equations shows that $\theta = 0$ lead us to the relation that we are looking
for {\it i.e.} \be C \propto  g^{2}_{k} \mu_{tc}, \label{eqC} \ee which give masses of $O(GeV)$. If the $SU(k)$
(or etc) bosons connect quarks to other ordinary fermions we also have \be A \propto  g^{2}_{k} \mu_{qcd},
\label{eqA} \ee which are masses of a few $MeV$. To obtain Eqs.(\ref{eqC}) and (\ref{eqA})  we neglected the
logarithmic term that appears in Eq.(\ref{eqfin2}). In principle there is no problem to assume the existence of
a tc dynamical self-energy with $\theta = 0$. There are tc models where it has been assumed that the self-energy
is dominated by higher order interactions that are relavant at or above the tc scale leading naturally to a very
hard dynamics \cite{carpenter,soni}. The existence of a hard self-energy in qcd is the unusual ingredient that
we are introducing here. Usually it is assumed that such solution is not allowed due to a standard operator
product expansion (OPE) argument \cite{politzer}. This argument does not hold if there are higher order
interactions in the theory or a nontrivial fixed point of the qcd (or tc) $\beta$ function \cite{holdom}. There
are many pros and cons in this problem which we will not repeat here \cite{lqcd}, but we just argue that several
recent calculations of the infrared qcd (or any non-Abelian theory) are showing the existence of an IR fixed
point \cite{alkofer} and the existence of a gluon (or technigluon) mass scale which naturally leads to an IR
fixed point \cite{ans}. The existence of such a mass scale seems to modify the structure of chiral symmetry
breaking  \cite{mns}.  This fact is not the only one that may lead to a failure of the standard OPE argument.
For instance, the effect of dimension two gluon condensates, if they exist, \cite{dim2} can also modify the
dynamics of chiral symmetry breaking and this possibility has not been investigated up to now. Therefore it
seems that we still do not have a full understanding of the IR behavior of the non-Abelian theories, which can
modify the behavior of the self-energies that we are dealing with. According to this we will just assume that
such behavior can happen in tc as well as in qcd. How much this is a bad or good assumption it will be certainly
reflected in the fermionic spectrum that we shall obtain. Finally, this is our only working hypothesis and will
lead us to the following problem: How can we prevent the coupling of the first and second fermionic generations
to the technifermions? A model along this line is proposed in the next section.
\section{The model}
\subsection{The fermionic content and couplings}
\begin{figure}[ht]
\begin{center}
\includegraphics{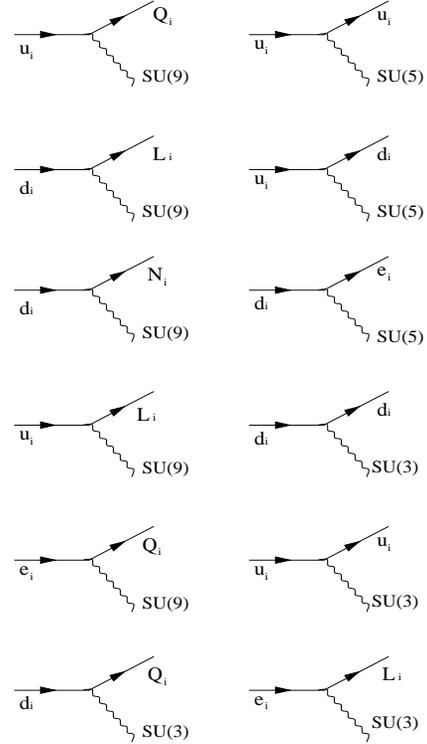} \vspace{0.5cm} \caption{Couplings of ordinary fermions and technifermions
to the gauge bosons of $SU(9)$, $SU(5)_{gg}$ and $SU(3)_{H}$ which are relevant for the generation of fermion
masses.} \label{fig1}
\end{center}
\end{figure}

\begin{figure}[ht]
\begin{center}
\includegraphics{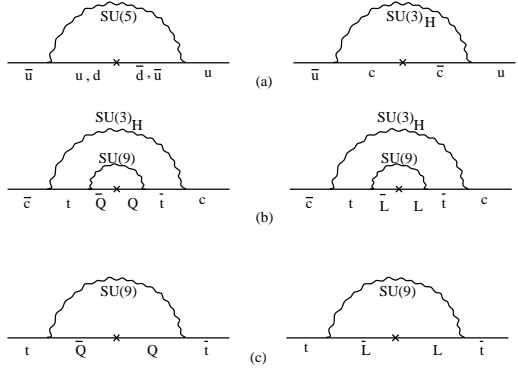}
\vspace{0.3cm} \caption{Diagrams contributing to the charge {2}/{3} quark masses. In (a) we indicate by $SU(5)$
the exchange of a boson that belongs to the $SU(9)$ group, but that would also appear in the minimal $SU(5)$
gut.} \label{f11}
\end{center}
\end{figure}

\begin{figure}[ht]
\begin{center}
\includegraphics{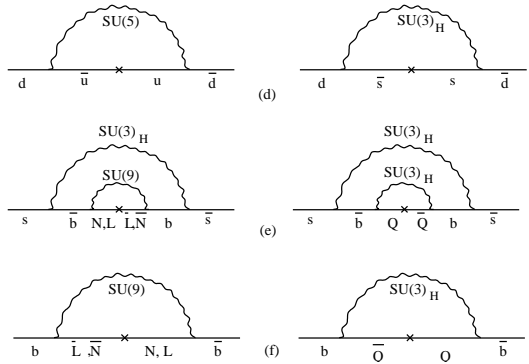} \vspace{0.3cm} \caption{Diagrams contributing to the mass generation of 1/3 charged quarks.}
\label{f12}
\end{center}
\end{figure}

\begin{figure}[ht]
\begin{center}
\includegraphics{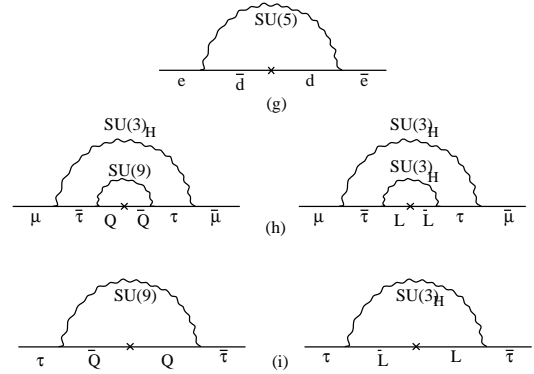} \vspace{0.3cm} \caption{Diagrams contributing to lepton masses.}
\label{f13}
\end{center}
\end{figure}

According to the dynamics that we proposed in the previous section, which consists in a self-energy with $\theta
=0$ in Eq.(\ref{mger1}), and as the different fermion masses will be generated due to the interaction with
different strong forces, we must introduce a horizontal (or family) symmetry to prevent the first and second
generation ordinary fermions to couple to technifermions at leading order. The lighter generations will couple
only to the qcd condensate or only at higher loop order in the case of the tc condensate. Using the hard
expression for the self-energy (Eq.(\ref{einc})) the fermion masses will depend only logarithmically on the
masses of the gauge bosons connecting  ordinary fermions to technifermions. Therefore we may choose a scale for
these interactions of the order of a gut scale, without the introduction of large changes in the value of the
fermion masses. We stress again that the only hypothesis introduced up to now is the dynamics described in the
previous section. On the other hand, as we shall see in the sequence, we will substitute the need of an extended
technicolor group by the existence of a quite expected unified theory containing tc and the standard model (SM)
at a gut scale. There is also another advantage in our scheme: It will be quite independent of the physics at
this ``unification" scale and will require only a symmetry (horizontal) preventing the leading order coupling of
the light fermion generations to technifermions.  Finally, the horizontal symmetry will be a local one, although
we expect that a global symmetry will also lead to the same results. We consider a unified theory based on the
$SU(9)$ gauge group, containing a $SU(4)_{\tiny{tc}}$ tc group (stronger than qcd) and the standard model, with
the following anomaly free fermionic representations \cite{ref1} \be 5\otimes[9,8] \oplus 1\otimes [9,2] \ee
\noindent where the $[\underline{8}]$ and $[\underline{2}]$ are \,antisymmetric \,under \,$SU(9)$. \,Therefore
\,the \,fermionic\, content \,of\, these\, representations\, can\, be\, decomposed \,according \,to\, the\,
\,group \,product \,\,$SU(4)_{{\tiny{tc}}}\otimes SU(5)_{{\tiny{gg}}}$ ($SU(5)_{\tiny{gg}}$\, is \,the\,
\,standard\, Georgi-Glashow \,gut
\cite{gg}) as: \br &&\hspace{-1cm}[\bf{9},\bf{2}]\nonumber\\
&&(1,10) = \left(\begin{array}{ccccc} 0 & \bar{u_{i}}_{B} & - \bar{u_{i}}_{Y} & -{u_{i}}_{R} & -{d_i}_{R}  \\
-\bar{u_i}_{B} & 0 & \bar{u_i}_{R} & -{u_i}_{Y} &  -{d_{i}}_{Y} \\ \bar{u_{i}}_{Y} & -\bar{u_{i}}_{R} & 0
& -{u_i}_{B} &  -{d_{i}}_{B} \\ {u_i}_{R} & {u_i}_{Y} & {u_i}_{B} & 0 & \bar{e_i}\\
{d_i}_{R} & {d_i}_{Y} & {d_i}_{B} & -\bar{e_{i}}  & 0\end{array}\right)\nonumber\\\nonumber \\ &&(4,5) =
\,\,\,\left(\begin{array}{c} {Q_i}_{R} \\ {Q_i}_{Y} \\ {Q_i}_{B} \\ \bar{L_i}\\ \bar{N_i}
\end{array}\right)_{TC}\,\,\,,\,\,\,(\bar{6},1)= N_{i}\nonumber \\ \nonumber \\
&&\hspace{-1cm}[\bf{9},\bf{8}]\nonumber\\
&&(1,\bar{5}) =\,\,\, \left(\begin{array}{c} \bar{d_i}_{R} \\ \bar{d_i}_{Y} \\ \bar{d_i}_{B} \\
e_i \\ \nu_{e_i}
\end{array}\right)
\,\,\,,\,\,\,(1,\bar{5}) = \left(\begin{array}{c} \bar{X}_{R_{k}} \\ \bar{X}_{Y_{k}} \\ \bar{X}_{B_{k}} \\
E_{k} \\ N_{E_{k}}
\end{array}\right)_i\nonumber\\ \nonumber \\\nonumber\\
&&(\bar{4},1)= \,\,\,\,\,\bar{Q_i}_{\varepsilon}, L_i ,{N_i}_{L} , \nonumber \er
\noindent where  $\varepsilon = 1..3$ is a color index  and
$k=1..4$ indicates the generation number of exotic fermions that must be introduced in order to render
the model anomaly free. These fermions will acquire masses of the order of the grand unified scale.
We are also indicating a generation (or horizontal) index $i=1..3$, that will appear due to the necessary
replication of families associated to a  $SU(3)_H$ horizontal group. This model is a variation of a model proposed
by one of us many years ago \cite{natale}.
The mass matrix of Eq.(\ref{e1}) will be formed according to the representations of the strongly interacting
fermions of the theory under the $SU(3)_H$ group. The technifermions form a quartet under  $SU(4)_{tc}$
and the quarks are triplets of qcd. The technicolor and color condensates will be formed at the scales
$\mu_{tc}$ and $\mu_{qcd}$ in the most attractive
channel (mac) \cite{suscor} of the products  ${\bf \bar{4}\otimes 4}$ and ${\bf \bar{3}\otimes 3}$ of each strongly interacting theory.
We assign the horizontal quantum numbers to technifermions and quarks such that these same products
can be decomposed in the following representations of  $SU(3)_H$:  ${\bf \overline{6}}$ in the case of the technicolor
condensate, and  ${\bf 3}$ in the case of the qcd condensate. For this it is enough that the standard
left-handed (right-handed) fermions transform as triplets (antitriplets) under $SU(3)_H$, assuming that the tc and qcd
condensates are formed in the ${\bf \overline{6}}$ and in the ${\bf 3}$ of  the $SU(3)_H$ group. This is consistent
with the mac hypothesis \cite{suscor} although a complete analysis of this problem is out
of the scope of this work.
The above choice for the condensation channels is crucial for our model, because the tc condensate in
the representation  ${\bf \overline{6}}$  (of $SU(3)_H$) will interact only with the third fermionic generation while the  ${\bf 3}$
(the qcd condensate) will interact
only with the first generation. In this way we can generate the coefficients $C$ and $A$ respectively of Eq.(\ref{e1}),
because when we add these condensates (vevs) and write them as a $3 \times 3$ matrix  we will end up ({\sl at leading order}) with
 \br
 M_f =\left(\begin{array}{ccc} 0 & A & 0\\ A^* & 0 & 0 \\
0 & 0 & C
\end{array}\right).
\label{e12} \er
This problem is very similar to the one proposed by Berezhiani and Gelmini {\it et al.} \cite{ref3} where the  vevs of fundamental
scalars are substituted by condensates.
The  new couplings generated by the unified $SU(9)$ group and by  the horizontal symmetry   $SU(3)_H$ are
shown in Fig.(\ref{fig1}).
With the couplings shown in Fig.(\ref{fig1}) we can determine the diagrams that are going to contribute to the
$2/3$ and $1/3$ charged quark masses as well as to the charged lepton masses. These diagrams are respectively
shown in Fig.(\ref{f11}) to (\ref{f13}). It is important to observe the following in the above figures:  The
second generation fermions obtain masses only at a two loop order. This mass will be proportional to $\mu_{tc}$
times two small couplings ($g_h$ and $g_9$, respectively the $SU(3)_H$ and $SU(9)$ coupling constants). It will
also be nondiagonal in the $SU(3)_H$ indices. The first generation fermions obtain masses only due to the qcd
condensate whereas the third generation ones couple directly to the tc condensates. Due to the particular choice
of representations under the unified theory containing tc and the standard model we end up with more than one
mass diagram for several fermions. It is particularly interesting the way the fermions of the first generation
obtain masses. In some of the diagrams of the above figures we show a boson that is indicated by $SU(5)$. This
boson belongs to the $SU(9)$ group, but would also appear in the standard $SU(5)_{\tiny{gg}}$ gut. For example,
the electron only couples to the $d$ quark (and to the qcd condensate) through a $SU(5)_{\tiny{gg}}$ gauge boson
existent in the Georgi-Glashow minimal gut, whereas the $u$ and $d$ can connect to the second generation through
the horizontal symmetry gauge bosons. We also expect other diagrams at higher order in $g_h$ and/or $g_9$ that
are not drawn in these figures.
\vspace{-0.1cm}
\vspace{-0.1cm}
\subsection{The composite Higgs system}
We can also observe that the second generation fermions will be massive not looking at the diagrams of
Fig.(\ref{f11}) to (\ref{f13}), but studying the composite Higgs system.  With this we mean that the qcd and tc
condensates act as if we had two composite bosons represented by the fields  $\eta$ and $\varphi$. In principle
this system could be described by the following effective potential \be
 V(\eta,\varphi) = \m^2_{\eta}\eta^{\dagger}\eta + \lambda_{\eta}(\eta^{\dagger}\eta)^2 +
\m^2_{\varphi}\varphi^{\dagger}\varphi + \lambda_{\varphi}(\varphi^{\dagger}\varphi)^2,
\label{vh}
\ee
in such a way that we can identify the vevs (given by the ratio of masses and couplings)
\be
v^{2}_{\eta}=-\frac{\m^2_{\eta}}{\lambda_{\eta}}\,\,\,,\,\,\,v^{2}_{\varphi}
=-\frac{\m^2_{\varphi}}{\lambda_{\varphi}} ,
\label{mc}
\ee
to the qcd and tc vacuum condensates.
The bosons  represented by $\eta$ and $\varphi$, respectively, are related to the system of
composite  Higgs bosons formed in the representations ${\bf 3}$ and $\overline{\bf {6}}$ of the horizontal group.
Such supposition is quite plausible if we consider the results of Ref.\cite{carpenter,soni}, where it was
shown that the interactions of a composite Higgs boson is very similar to the ones of a fundamental boson.
Our intention is to show that such system leads to an intermediate mass scale and to a mass matrix identical
to Eq.(\ref{e1}).
\par The vevs of qcd and technicolor, due to the horizontal symmetry, can be written
respectively in the following form \cite{ref3}
\be
 \langle\eta\rangle \sim \left(\begin{array}{c} 0 \\ 0 \\  v_{\eta}
\end{array}\right)\,\, \,,\,\,\, \langle\varphi\rangle \sim \left(\begin{array}{ccc} 0 & 0 & 0\\ 0 & 0 & 0 \\ 0 & 0
& v_{\varphi} \end{array}\right),
\label{veta}
\ee
and will be of the order of   $250$ MeV and $250$ GeV.
It is instructive at this point to observe what fermionic mass matrix we can obtain with the
vevs of Eq.(\ref{veta}). We can assume that the composite scalars $\eta$ and
$\varphi$ have ordinary Yukawa couplings \cite{fritzsch,ref3} to fermions described by the
following effective Yukawa lagrangian
\be {\cal{L}}_{Y} = a\bar{\Psi}^{i}_{L\lambda}\eta^{k}_{\lambda}U^{j}_{R}\epsilon_{ijk} +
b\bar{\Psi}^{i}_{L\lambda}\varphi^{ij}U^{j}_{R} ,
\label{y1} \ee
where  $\Psi$ and $U$ are the ordinary fermion fields. $\lambda$ is a weak hypercharge ($SU(2)_{w}$) index,
for instance, $\lambda = 1$ represents charge $2/3$ quarks and $\lambda  = 2$ correspond to the charge $1/3$
quarks, $i,j$ e $k$ indicate the components of the composite scalar bosons of
the representations ${\bf 3}$ and $\overline{\bf {6}}$ of $SU(3)_{H}$
and $a$ and $b$ are the coupling constants.
Substituting the vevs of Eq.(\ref{veta}) in the Yukawa lagrangian for the charge $2/3$ quarks, we obtain
\be {\cal{L}}_{Y} = a\bar{c}_{L}v_{\eta}u_{R} - a\bar{u}_{L}v_\eta{c}_{R} + b\bar{t}_{L}v_{\varphi}t_{R} ,
\label{yukl}
\ee
leading to a mass matrix in the $(u\,,c\,,t)$ basis which is given by
\be \overline{m}^{\frac{2}{3}} = \left(\begin{array}{ccc} 0 & -av_{\eta} & 0\\ av_{\eta} & 0 & 0 \\ 0 & 0 & bv_{\varphi}
\end{array}\right). \ee
The main point of the model is that the fermions of the third generation obtain large masses because they couple
directly to technifermions, while the ones of the first generation obtain masses originated by the ordinary
condensation of qcd quarks.  Having this picture on mind we can now see that the most general vev for this
system includes the mass generation for the intermediate family.
\par
\begin{figure}[ht]
\begin{center}
\includegraphics{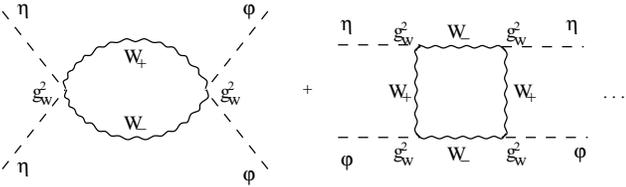} \caption{Higher order corrections coupling the $\eta$ and
$\varphi$ composite bosons.} \label{f3}
\end{center}
\end{figure}

 It is important to verify that there is no way to prevent the coupling at higher order of
the different composite scalar bosons with $SU(3)_H$ quantum numbers. Examples of such couplings are shown in Fig.(\ref{f3})
\par
\begin{figure}[ht]
\begin{center}
\includegraphics{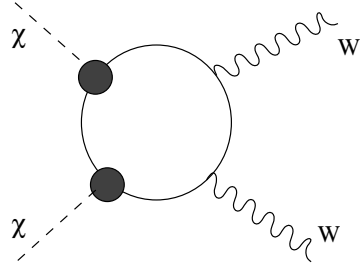} \caption{Diagram leading to the coupling between two composite scalar bosons and two gauge
bosons} \label{f6}
\end{center}
\end{figure}

\begin{figure}[ht]
\begin{center}
\includegraphics{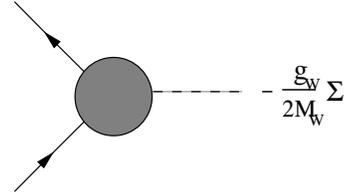} \caption{Vertex coupling a scalar composite boson to ordinary fermions} \label{f5}
\end{center}
\end{figure}

 The diagrams of Fig.(\ref{f3}) will produce new terms for the effective potential of our composite
 system, therefore we must add to Eq.(\ref{vh}) the following terms
\be V_{2}(\eta,\varphi) =  \Pi\eta^{\dagger}\eta\varphi^{\dagger}\varphi +
\delta\eta^{\dagger}\varphi\eta\varphi^{\dagger} + ... \label{vh2} \ee The introduction of this expression in
the potential of Eq.(\ref{vh}) will shift the vevs generated by the effective fields $\eta$ and  $\varphi$, and
the vev associated to the field $\eta$ will be shifted to \be \langle\eta\rangle \sim \left(\begin{array}{c} \varepsilon \\
0
\\  v_{\eta}
\end{array}\right) .
\label{veshi}
\ee
\noindent We do not include the shift in the vev of $\varphi$, because $v_{\eta}\ll v_{\varphi}$
and the modification is negligible.
Note that the Yukawa lagrangian that we discussed in Eq.(\ref{yukl}) in terms of the new vevs can
be written as
\be {\cal{L}}_{Y} = \,\, a\bar{c}_{L}v_{\eta}u_{R} - a\bar{u}_{L}v_\eta{c}_{R} + b\bar{t}_{L}v_{\varphi}t_{R} - a\bar{c}_{L}\varepsilon t_{R} +
a\bar{t}_{L}\varepsilon c_{R}.
 \ee
Therefore, in the $(u\,,c\,,t)$ basis, the structure of the mass matrix now is
\be \overline{m}^{\frac{2}{3}} = \left(\begin{array}{ccc} 0 & -av_{\eta} & 0\\ av_{\eta} & 0 & a\varepsilon \\ 0 &
-a\varepsilon & bv_{\varphi}
\end{array}\right) . \ee
This example was motivated by a system of fundamental Higgs bosons \cite{ref3}. But the most remarkable fact is that
we can reproduce this result with a composite system formed by the effective low energy theories coming from qcd and tc as
we shall see in the following.
The coefficient  $\varepsilon$ in Eq.(\ref{veshi}) will result from the minimization of the full potential
\br V(\eta,\varphi) = \,\,&&\m^2_{\eta}\eta^{\dagger}\eta + \lambda_{\eta}(\eta^{\dagger}\eta)^2 +
\m^2_{\varphi}\varphi^{\dagger}\varphi + \lambda_{\varphi}(\varphi^{\dagger}\varphi)^2 + \nonumber\\
&&\Pi\eta^{\dagger}\eta\varphi^{\dagger}\varphi + \delta\eta^{\dagger}\varphi\eta\varphi^{\dagger}.
\label{vcalc} \er This coefficient can be calculated if we assume that   $\langle\eta\rangle$ is given by
Eq.(\ref{veshi}), $\langle\varphi\rangle$ is the same vev described by Eq.(\ref{veta}) and both are related to
the tc and qcd condensates. We will neglect   $\delta$ compared to $\Pi$ in Eq.(\ref{vh2}), what is reasonable
if we look at Fig.(\ref{f3}) ($\Pi$ is given by the first diagram). The coupling $\Pi$ is computed from the
first diagram of Fig.(\ref{f3}) using the effective vertex   $\chi \chi  WW$ shown in Fig.(\ref{f6}), where an
ordinary fermion runs in the loop,
\noindent where the $\chi$ field may indicate technicolor  ($\chi = \eta $) or qcd  ($\chi = \varphi$) composites scalar bosons.
To compute Fig.(\ref{f6}) we also need the effective coupling between the composite scalars boson and the ordinary
fermions. This one has been calculated in the work of Carpenter et al. \cite{carpenter,soni} some
years ago and it is shown in Fig.(\ref{f5}).
After a series of steps the calculation of the diagram of Fig.(\ref{f6}) will be given by \be \Pi_{\chi\chi{WW}}
\sim -\frac{g^4_{W}\delta^{ab}}{M^2_{W}}\frac{g^{\m\nu}}{32\pi^2}\int\!d^2q\frac{\Sigma^2_\chi}{q^2}.
\label{c22} \ee \noindent Following closely the procedure adopted by Carpenter et al. \cite{soni} we may
approximate the self energy by $\Sigma_\chi\! \sim \!\m_{\chi}
\!\left(\frac{q^2}{\m_\chi^2}\right)^{-\varsigma}\!\!\!\!\!\!\!,$ where
$\varsigma_\chi=\frac{3C_{2\chi}g^2_{\chi}}{16\pi^2}$ , to obtain the following coupling between two composite
scalars and the intermediate gauge bosons of the weak interaction \be \Pi_{\chi\chi{WW}} \sim
-\frac{M^2_{W}\delta^{ab}}{2\pi^2}\frac{G^{2}_{F}\m^2_{\chi}}{\varsigma_{\chi}}g^{\m\nu}\!\!. \label{cf22} \ee
\noindent \noindent In Eq.(\ref{cf22})  we made use of the relation
$\frac{G_{F}}{\sqrt{2}}=\frac{g^2_{W}}{8M^{2}_{W}}$. Note that the coupling between scalars and gauge bosons is
dominated by the ultraviolet limit, where the approximation for the self energy discussed above is also valid.
The effective coupling $\Pi$ in Eq.(\ref{vcalc}) is equivalent to the calculation of the first diagram of
Fig.(\ref{f3}). Using Eq.(\ref{cf22}) we will come to the following expression \be \Pi_{\eta\eta\varphi\varphi}
= \frac{M^4_{W}{G^4_{F}}\m^2_{tc}\m^2_{qcd}}{32\pi^8\varsigma_{tc} \varsigma_{qcd}}. \label{pia} \ee
\par We can now approximately determine the value of $\varepsilon$ assuming that the potential
of Eq.(\ref{vh}) has a minimum described by the vevs $\langle\varphi\rangle$, Eq.(\ref{veta}), and
$\langle\eta\rangle$, Eq.(\ref{veshi}), what lead us to  the following value of the potential at minimum \be
V(\eta,\varphi)|_{min}=\m^2_{\eta}v^2_{\eta}+ \lambda_{\eta}v^4_{\eta} + \m^2_{\varphi}v^2_{\varphi}+
\lambda_{\varphi}v^4_{\varphi} + \lambda_{\eta}\varepsilon^4. \ee We then compare the minimum of this potential
with the one obtained from Eq.(\ref{vcalc}), where the term proportional to $\delta$ is neglected in comparison
to the one proportional to  $\Pi$. This is equivalent to say that the second diagram of Fig.(\ref{f3}) is much
smaller then the first diagram, and the vevs entering in  Eq.(\ref{vcalc}) are the unperturbed ones because the
perturbation will enter through the  $\Pi$ term. Finally, assuming that the coefficient describing the coupling
between four scalar bosons that are formed in the chiral symmetry breaking of QCD is given by\cite{soni} \be
\lambda_{\eta} = \frac{G^2_{F}\m^4_{qcd}c_{qcd}\alpha_{qcd}}{\pi}, \ee and we can obtain a similar expression
for  $\lambda_{\varphi}$ after changing the indices $qcd$ by $tc$. Equalizing $v_{\eta}$ and $v_{\varphi}$ to
the known qcd and tc condensates (assuming  $ \< \bar{\psi}_{i}\psi_{i} \> = v^3 \approx \mu_i^3$ \cite{hg}), we
conclude that \be \varepsilon \sim  B  \sim
\left(\frac{M^4_{W}G^2_{F}\m^4_{tc}}{18\pi^3c_{tc}\alpha_{tc}}\right)^{\frac{1}{4}}GeV \sim 16.8 GeV.
\label{mc12} \ee The surprising fact in this calculation is that the coupling of the different scalar bosons has
been determined dynamically and gives exactly the expected value for the nondiagonal coefficient $B$. In models
with fundamental scalar bosons this value results from one  {\sl ad hoc} choice. In this section we presented
our model, determined the main diagrams contributing to the fermion masses and showed that this scenario
naturally leads to a fermion mass matrix with the Fritzsch texture. We have not tested many other models, but it
seems that we may have a full class of models along the line that we are proposing here.  Because of the
peculiar dynamics that we are assuming we need only a horizontal symmetry and a partial unification of the
standard model and the value of their mass scales will not strongly modify our predictions (although the chosen
horizontal symmetry will). Of course, the breaking of the unified and/or horizontal symmetry will happen at a
very high energy scale and will not be discussed here. In particular, this symmetry breaking can be even
promoted by fundamental scalars which naturally  can appear near the Planck scale.
\section{Computing the mass matrix}
We can now compute the mass matrix. Let us first consider only the $\frac{2}{3}$ charged quarks and verify their different contributions
to the matrix in Eq.(\ref{e1}). These will come from the diagrams labeled (a), (b)  and (c)  in
Fig.(\ref{f11}) and are equal to
\br A\,\,=\,\,&&\frac{\mu_{qcd}}{10 c_{qcd}\alpha_{qcd}}\left[ 1 +
bg^2_{qcd}ln\frac{M^2_{5}}{\m^2_{qcd}}\right]^{-\gamma_{qcd} + 1} \hspace{-1cm} +  \nonumber\\
&&\frac{4\mu_{qcd}}{135 c_{qcd}\alpha_{qcd}}\left[ 1 + bg^2_{qcd}ln\frac{M^2_{h}}{\m^2_{qcd}}\right]^{-\gamma_{qcd} +
1}\nonumber\hspace{-1.2cm},\\\nonumber\\
B\,\,=\,\,&& \frac{28\mu_{tc}}{675\pi c_{tc}\alpha_{tc}}\left[ 1 +
bg^2_{tc}ln\frac{M^2_{9}}{\m^2_{tc}}\right]^{-\gamma_{tc} + 1}\nonumber\hspace{-1.0cm},\\\nonumber\\
C\,\,=\,\,&&\,\,\,\,\frac{2\mu_{tc}}{15 c_{tc}\alpha_{tc}}\left[ 1 +
bg^2_{tc}ln\frac{M^2_{9}}{\m^2_{tc}}\right]^{-\gamma_{tc} + 1}\hspace{-1cm}. \label{abc} \er Where the
contributions for $A$, $B$ and $C$ come respectively from the diagrams (a), (b)  and (c) displayed in
Fig.(\ref{f11}). The values  $A$, $B$ and $C$ correspond to the nondiagonal masses in the horizontal
symmetry basis. To come to these values we assumed $\alpha_{k} (= \alpha_9 =\alpha_h =\alpha_5) \sim \frac{1}{45}$ at the
unification scale. We also assumed, when computing diagrams involving the technileptons and techniquarks
condensates, the following relation
\be
 \langle\bar{L} { L\rangle} = \frac{ 1}{ 3}{ \langle\bar{Q} Q\rangle}  ,
\ee because the techniquarks carry also the three color degrees of freedom. As the mass matrix is the same
obtained in Ref.\cite{fritzsch} we can use the same diagonalization procedure to obtain the $t$, $c$ and $u$
quark masses, which is given by \be M^{\frac{2}{3}}_{f_{Diag}} = R^{-1}M^{\frac{2}{3}}_{f}R, \ee where $R$ is a
rotation matrix described in Ref.\cite{fritzsch}. After diagonalization we obtain \be m_{u}\, \sim\, \frac{{\mid
A\mid}^2}{{\mid B\mid}^2}\mid C\mid\,\,,\,\, m_{c}\, \sim\, \frac{{\mid B\mid}^2}{\mid C\mid}\,\,\,\,{\rm
and}\,\,\,\,m_{t}\,\sim\, \,\,\mid C\mid, \ee \noindent where  the values of  $A$, $B$ and $C$ are the ones
shown in Eq.(\ref{abc}).  We will also assume the unification mass scale as $M_9 = M_5 \sim 10^{16}$ GeV and the
horizontal mass scale equal to  $M_{h}\sim 10^{13}$ GeV. The several constants contained in Eq.(\ref{abc})  are
$b_{tc}= \frac{1}{16\pi^2}\frac{26}{3}$, $b_{qcd}= \frac{7}{16\pi^2}$, $\gamma_{tc}=\frac{15}{23}$ and $
\gamma_{qcd}=\frac{4}{7}$. We remember again that we assumed $\alpha_{k} \sim \frac{1}{45}$,  $\m_{tc} = 250
GeV$ and $\m_{qcd} = 250 MeV$. The  fermion masses  come out as a function of the  parameter $c_{(tc\,,\,qcd)}
\alpha_{(tc\,,\,qcd)}$. For simplicity (as well as a reasonable choice) we will define   $c\alpha = c_{tc}\alpha
_{tc} = c_{qcd}\alpha_{qcd}=0.5$.
\par We display in Table 1 the fermionic mass spectrum obtained in this model.
Some of the values show a larger disagreement in comparison to the experimental values,  and others show a quite
reasonable agreement if we consider all the approximations that we have performed and the fact that we have a
totally dynamical scheme.
\begin{table}[h]
\begin{center}
\begin{tabular}{|c|c|c|c|c|c|}
  \hline
   $m_{t}$  &  160.3 GeV  &   $m_{b}$    &   113 GeV   &  $m_{\tau}$ &  131.2 GeV   \\
   $m_{c}$  & 1.57 GeV    &   $m_{s}$    &  1.10 GeV   &  $m_{\mu}$  &   1.30 GeV  \\
   $m_{u}$  &  29.6 Mev   &   $m_{d}$    &  15.6 Mev   &  $m_{e}$    &    5.5 Mev  \\
   \hline
\end{tabular}
 \end{center}
\caption{Approximate values for  quarks and leptons  masses according to the chosen values of couplings and
strongly interacting mass scales.}
\end{table}
\par It is also impressive that $B$ in Eq.(\ref{abc}), neglecting logarithimic terms, is roughly giving by  $B
\sim 14 \alpha_h m_t / \pi$  which is of order of $17$ GeV. This is the expected value according to the
estimative of the previous section (see Eq.(\ref{mc12})). In some way this is also expected in a mechanism where
one fermionic generation obtain a mass at 1-loop level coupling to the next higher generation fermion (see, for
instance, Ref.\cite{ref5}). The values of the $u$ and $e$ masses can be easily lowered with a smaller value of
$\mu_{qcd}$. Of course, we are also assuming a very particular form for the mass matrix based in one particular
family symmetry. Better knowledge of the symmetry behind the mass matrix, and a better understanding of the
strong interaction group alignment will certainly improve the comparisom between data and theory. The high value
for the masses obtained for some of the second generation fermions also come out from the overestimation of the
$b$ and $\tau$ masses. The mass splitting between the $t$ and $b$ quarks, which is far from the desirable
result, is a problem that has not been satisfactorily solved in most of the dynamical models of mass generation
up to now. It is possible that an extra symmetry, preventing these fermions to obtain masses at the leading
order as suggested by Raby \cite{raby} can be easily implemented in this model.  We will discuss these points
again in the conclusions. Finally considering that we do not have any flavor changing neutral current problems
\cite{dime}, because the interaction between fermion and technifermions has been pushed to very high energies,
and that we assume only the existence of quite expected symmetries  (a gauge group containing tc and the
standard model and a horizontal symmetry) the model does quite well in comparison with many other models.
\section{Pseudo-Goldstone boson masses}
\par  Another problem in technicolor models is the proliferation of pseudo-Goldstone bosons \cite{weisus,hillsi,ref6}.
After the chiral symmetry breaking of the strongly interacting sector a large number of Goldstone bosons
are formed, and only few of these degrees of freedom are absorbed by the weak interaction gauge bosons. The others
may acquire small masses resulting in light  pseudo-Goldstone bosons that have not been observed experimentally.
In our model these bosons obtain masses that are large enough to have escaped detection
at the present accelerator energies, but will show up at the next generation of accelerators (for instance, LHC).
We can list the possible pseudo Goldstone bosons according to their different quantum numbers:
\par \textit{ Colored pseudos}: They carry color degrees of freedom and can be divided into the
{\bf{3}} or {\bf{8}} color representations. We can indicate them by $$\Pi^a \sim \bar{Q}\gamma_5\lambda^aQ. $$
\par \textit{ Charged pseudos}: These ones carry electric charge and we can take as one example the following
current $$\Pi^{+}\sim \bar{L}\gamma_5Q,$$
\noindent where $Q(L)$ indicate the techniquark (technilepton) fields.
\par \textit{ Neutral pseudos}: They do not carry color or charge and one example is
$$\Pi^0 \sim \bar{N}\gamma_5N.$$
\par Following closely Ref.\cite{ref6} the standard procedure to determine the  $SU(3)_{qcd}$
contribution to the mass ($M_c$) of a colored pseudo Goldstone boson gives
\br M_{c}\,&&\sim \left(\frac{C_2(R)\alpha_{c}(\mu)}{\alpha_{el}}\right)^{\frac{1}{2}}\frac{F_\Pi}{f_\pi}35.5
MeV \nonumber\\&&\sim 170\sqrt{C_2(R)} GeV \sim O(300)GeV. \er
\par While the electromagnetic contribution to the mass of the charged pseudos Goldstone bosons is
estimated to be \cite{ref6}
\be
M_{em}\sim Q_{ps}\frac{F_{\Pi}}{f_{\pi}}35.5 MeV \sim Q_{ps}47 GeV \sim O(50 GeV),
\ee
 \noindent in the equations above we assumed that the technipion and pion decay constants
 are given by $F_{\Pi}\approx 125 GeV$ and  $f_{\pi}\approx 95 MeV$, $Q_{ps}$ is the electric
 charge of the pseudo-Goldstone boson, and  $C_2(R)$ is the quadratic Casimir operator in the representation
 $R$ of the pseudo-Goldstone boson under the tc group. There is not much to change in these standard
 calculations, except that due to the particular form of the technifermion self energy the technifermion will
 acquire large current masses, and subsequently the  pseudos-Goldstone bosons formed with these ones.
We know that any chiral current
 $\Pi^f$ can be written as a vacuum term $m_f  \langle\bar{\psi}_f\psi_f\rangle $ plus electroweak (color, ...)
 corrections \cite{adriano}, where $m_f$ is the current mass of the fermion $\psi_f$ participating
 in the composition of the current  $\Pi^f$, neglecting the electroweak corrections and
 using PCAC in the case of qcd we obtain the Dashen relation
\be m^2_{\pi} \approx \frac{m_{q}\langle\bar{q}q\rangle}{f^2_{\pi}}, \ee \noindent where
$\langle\bar{q}q\rangle$ is the quark condensate. Of course this relation is valid for any chiral current and in
particular for the technifermions we can write \be M^2_{\Pi} \approx
\frac{M_{T_{f}}\langle\bar{T_{f}}T_{f}\rangle}{F^2_{\Pi}}, \label{MPG} \ee \noindent where $M_{T_{f}}$ is the
technifermion current mass. In the usual models (with the self-energy given by Eq.(\ref{eq2})) the
technifermions are massless or acquire very tiny masses  leading to negligible values for $M_{\Pi}$. In our
model this is not true. All technifermions acquire masses due to the self-interaction with their own condensates
through the interchange of $SU(9)$ bosons.
\par
\begin{figure}[ht]
\begin{center}
\includegraphics{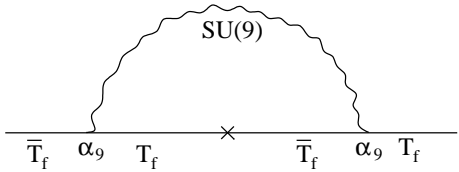} \caption{Diagram responsible for the technifermion mass generation.} \label{fx}
\end{center}
\end{figure}

 There are several bosons in the $SU(9)$ (and also in the $SU(3)_H$) theory connecting to technifermions
and generating a current mass as is shown in Fig.(\ref{fx}).
\par A simple estimative, based on  Eq.(\ref{einc}), of the contribution of Fig.(\ref{fx}) to the technifermion
masses  gives
\be
M_{T_{f_{SU(9)}}}\gsim O(80 - 130)GeV .
\ee
\noindent  If we also include the contribution of the same diagram where the exchanged boson
is a horizontal  $SU(3)_{H}$ boson coupling technifermions of different generations,
we must add to the above value the following one
\be
M_{T_{f_{SU(3)_{H}}}} \gsim O(10 - 40) GeV .
\ee
\par Therefore, we expect that the technifermion current masses are at least of the order of
$M_{T_{f}}\approx O(100)GeV$. Now, according to Eq.(\ref{MPG}) and assuming $\langle\bar{T_{f}}T_{f}\rangle\sim
F^3_{\Pi}$ we have the following estimative for the pseudo-Goldstone boson masses \be M_{\Pi}\gsim O(100) GeV .
\ee
\par Note that in this calculation we have not considered the qcd or electroweak corrections
discussed previously. Therefore, even if the pseudo-Goldstones bosons do not acquire masses
due to qcd or electroweak corrections they will at least have masses of order of  $100$ GeV
because of the ``current" technifermion masses obtained at the $SU(9)$ (or $SU(3)_H$) level.
\section{Conclusions}
We have presented a technicolor theory based on the group structure
 $SU(9) \times SU(3)_H$. The model is based on a particular ansatz
 for the tc and qcd self energy. We argue that our ansatz for qcd, in view of the many
 recent results about its infrared behavior, is a plausible one,  but even if
 it is considered as an ``ad-hoc" choice for the self energy the main point is
 that it leads to a consistent model for fermion masses. This is the
 only new ingredient in the model, all the others (unification of tc and the standard
 model and the existence of a horizontal symmetry)
 are naturally expected in the current scenario of particle physics.
 One of the characteristics of the model is that the first fermionic generation basically obtain
 masses due to the interaction with the qcd condensate, whereas the third generation
 obtain masses due to its coupling with the tc condensate. The reason for this particular
 coupling and for the alignment of the strong theory sectors generating intermediate masses is
 provided by the $SU(3)_H$  horizontal symmetry. Of course, our model is not successful in predicting
  all the fermion masses although it has a series of advantages. It does not need
  the presence of many etc boson masses to generate the different fermionic mass
  scales. The etc theory is replaced by an unified and horizontal symmetries. It has no
  flavor changing neutral currents or unwanted light pseudo-Goldstone bosons.
  There are many points that still need some work in this line of model. The breaking
  of the $SU(9)$ and horizontal symmetries is not discussed, and just assumed to happen
  near the Planck scale and possibly could be promoted by fundamental scalar bosons.
  The mass splitting in the third generation could be produced with the introduction of
  a new symmetry. For instance, if in the $SU(9)$ breaking besides the standard model
  interactions and the tc one we leave an extra $U(1)$, maybe we could have quantum numbers
  such that only the top quark would be allowed to couple to the tc condensate at leading order.
  This possibility should be further studied because it also may introduce large quantum corrections
  in the model. If the unified group ($SU(9)$ in our case) is not broken by a dynamical mechanism,
  {\it i.e.} we do not need that this group tumbles down to $SU(4)_{tc}\otimes SM$, then we could
  replace $SU(4)_{tc}$ by one smaller group (perhaps  $SU(2)_{tc}$) which becomes stronger
  at the scale $\mu \approx 250$ GeV.
  In this class of models we can choose different groups containing tc and the standard model,
  as well as different horizontal symmetries with
  different textures for the mass matrix.  These will certainly modify the values of the fermion
  masses that we have obtained. The alignment of the strongly interacting sectors can be
  studied only with many approximations, but it is quite possible that it generates more entries
  to the mass matrix than only the term $B$. Another great advantage of the model is that it
  is quite independent of the very high energy interactions (like $SU(9)$ or $SU(3)_H$), although
  the horizontal symmetry is fundamental to obtain the desired mass matrices, and we believe
  that variations of this model can be formulated.
\section{Acknowledgments}
This research was supported by the Conselho Nacional de Desenvolvimento
Cient\'{\i}fico e Tecnol\'ogico (CNPq) (AAN) and by Fundac\~ao de Amparo \`a
Pesquisa do Estado de S\~ao Paulo (FAPESP) (AD).

\end{document}